\documentclass{pasj00}
\begin{document}
\SetRunningHead{Naito et al.}{Discovery of Metastable Helium Absorption Lines in V1280 Sco}
%\Received{}%{yyyy/mm/dd}
%\Accepted{}%{yyyy/mm/dd}
%\Published{}%{yyyy/mm/dd}

\title{Discovery of Metastable Helium Absorption Lines in V1280 Scorpii\thanks{Based on data collected at the Subaru Telescope, which is operated by the National Astronomical Observatory of Japan.}}

%%% begin:list of authors
% Do NOT capitalize all letters in "textsc".
\author{Hiroyuki \textsc{Naito}} %
%  \thanks{Example: Present Address is xxxxxxxxxx}}
\affil{Graduate School of Science, Nagoya University, Furo-cho, Chikusa-ku, Nagoya 464-8602}
\email{naito@stelab.nagoya-u.ac.jp}

\author{Akito \textsc{Tajitsu}}
\affil{Subaru Telescope, National Astronomical Observatory of Japan, 650 North A'ohoku Place, Hilo, HI 96720, USA}

\author{Akira {\sc Arai}}
\affil{Center for Astronomy, University of Hyogo, Sayo-cho, Sayo, Hyogo 679-5313}
\and
\author{Kozo {\sc Sadakane}}
\affil{Astronomical Institute, Osaka Kyoiku University, Asahigaoka, Kashiwara, Osaka 582-8582}
%%% end:list of authors

%%% Please use the following style in case that sorting by 
%%% affiliation is impossible. 
%
% \author{%
%   D-Firstname \textsc{D-Familyname}\altaffilmark{1}
%   E-Firstname \textsc{E-Familyname}\altaffilmark{1,2}
%   and
%   F-Firstname \textsc{F-Familyname}\altaffilmark{2}}
% \altaffiltext{1}{Address of Institute}
% \email{ddddd@xxx.xxx.xx.xx}
% \email{eeeee@xxx.xxx.xx.xx}
% \altaffiltext{2}{Address of Institute}

%% `\KeyWords{}' always has to be placed before `\maketitle'.
 %Do NOT move this preamble from here!

\maketitle

\begin{abstract}
We report the discovery of blue-shifted metastable He~{\sc i*} absorption lines at 3188 \AA~ and 3889 \AA~ with multiple components on high-resolution spectra ($R$ $\sim$ 60\,000) of V1280 Scorpii. Similar multiple absorption lines associated with Na~{\sc i} D doublet and Ca~{\sc ii} H and K are observed. Na~{\sc i} D doublet absorption lines have been observed since 2009, while the metastable He~{\sc i*} absorption lines were absent in 2009 and were detected in 2011 (four years after the burst). We find different time variations in depths and velocities of blue-shifted absorption components among He~{\sc i*}, Na~{\sc i}, and Ca~{\sc ii}. The complex time evolutions of these lines can be explained by assuming changes in density and  recombination/ionization rate when the ejecta expand and the photosphere contracts to become hotter. The multiple absorption lines originate in the ejected materials consisting of clumpy components, which obscure a significant part of the continuum emitting region. We estimate the total mass of the ejected material to be on the order of $\sim$ 10$^{-4}$ M$_{\odot}$, using metastable He~{\sc i*} $\lambda$$\lambda$3188 and 3889 absorption lines.
\end{abstract}

%\noindent IMPORTANT NOTICE\\
%1. ``\verb|\draft|'' creates single column and double spaces format.\\
%2. If you comment out ``\verb|\draft|'', the output will be double column
%   and single space.\\
%3. For cross-references, the use of ``\verb|\label|, \verb|\ref|, \verb|\cite|'' 
%   and the thebibliography environment is strongly recommended. \\
%4. Do NOT use ``\verb|\def|, \verb|\renewcommand|''.\\
%5. Do NOT redefine commands provided by PASJ00.cls.\\

%\cite{newt1687} → Newton 1687
%\citep{newt1687} → (Newton 1687)
%\citet{newt1687} → Newton (1687)
%\authorcite{newt1687} → Newton
%\yearcite{newt1687} → 1687
%Fe~{\sc ii}

\section{Introduction}

V1280 Sco (Nova Scorpii 2007 No. 1) is a classical nova which was independently discovered by two Japanese amateur astronomers (Y. Nakamura and Y. Sakurai)  on 2007 February 4 at ninth visual magnitude \citep{yamaoka07}. Detailed descriptions of photometric and spectroscopic observations from pre-maxium to plateau phase  (from February 2007 to April 2011) have been reported in \citet{naito12}. Recently, \citet{chesneau12} reported the presence of a dusty hourglass-shaped bipolar nebula around V1280 Sco based on mid-infrared imaging observations.
%After the discovery, the nova had shown very strange changes in brightness, which underwent the initial rising phase ($\sim$10 days), the pre-maximum halt (2 days), three fluctuations ($\Delta$ V $\sim$ 1 mag $<$ 1 day for each fluctuation; \cite{hounsell10}) near the maximum light ($V_\mathrm{max}$ = 3.78; \cite{munari07}) and then rapid fading (followed by gradual recovering which began after $\sim$150 days the maximum) associated with dust formation.

A particularly interesting feature of this nova is that it had a exceptionally long plateau spanning over 1500 days, which suggests that V1280 Sco is an extremely slow nova. According to our latest photometric observation in the $y$ band at Osaka Kyoiku University, the nova maintains its brightness ($m_{y}$ = 10.45 $\pm$ 0.01) at least until 2012 August 2 ($\sim$ 66 months after the maximum light). %Spectral observations in both optical and near-IR regions were also performed, showing that V1280 Sco was an Fe~{\sc ii} type (dust-forming) nova and had a continuum-dominant spectrum accompanied by only low excited permitted lines (H~{\sc i}, He~{\sc i*} and Fe~{\sc ii}) and forbidden lines ([O~{\sc iii}] and [N~{\sc ii}]) $\sim$1600 days after the burst.
The nova took about 50 months after the burst to enter the nebular phase, which implies that V1280 Sco is the slowest nova in the historical record.

The very long evolutionary time-scale observed in V1280 Sco is in agreement with predictions of slow novae in the optically thick wind model \citep{kato94}. The model shows that the decreasing mass loss rate of the nova wind results in an inward moving photosphere and increase in the effective temperature and degree of ionization. In theory, based on the universal decline law propounded by \citet{hachisu06}, the speed class of classical novae is mainly determined by the white dwarf (WD) mass; progenitors of  slow novae are expected to be lower-mass WDs than fast novae. If the theory is the case, similar burst mechanisms work between slow and fast novae and there could be common (but on a different time scale) features in observational data. It is very important to seek common and/or missing observational features among classical novae to understand the nova phenomenon as a whole. In this respect, V1280 Sco is the best target because it is an extremely slow nova and provides us with a valuable opportunity to investigate its spectral evolution closely.

\citet{sadakane10} (hereafter, Paper I) discovered multiple high-velocity narrow Na~{\sc i} D absorption lines on high-resolution spectra obtained in 2009. They suggest that these high-velocity components originated in cool clumpy gas clouds moving on the line-of-sight, produced during interactions between pre-existing cool circumstellar gas and high-velocity gas ejected in the nova explosion. These absorption are useful to investigate the physical conditions and time evolution of surrounding circumbinary gas. Similar features of Na~{\sc i} D doublet and Ca~{\sc ii} H and K lines are observed in many other classical novae during the early phase \citep{williams08}, while the survival time (several years) of Na~{\sc i} D absorption lines in V1280 Sco is much longer than those observed among fast novae (within several weeks or months). To investigate the difference on multiple absorption lines between V1280 Sco (slow nova) and fast novae, we have continued high-resolution spectroscopic observations for four years (2009 -- 2012). 

In this paper, we report the discovery of blue-shifted metastable He~{\sc i*} absorption lines and investigate time variations of He~{\sc i*}, Na~{\sc i} D and Ca~{\sc ii} lines from 2009 to 2012.
Our spectral observations are summarized in section 2. We focus on the discovery of multiple metastable He~{\sc i*} absorption lines in section 3, and co-existance of He~{\sc i*}, Na~{\sc i} and Ca~{\sc ii} lines are described in section 4. We discuss the origins of discrete absorption lines and the structure of ejected shell in section 5, and conclusions are given in section 6.

%\begin{figure}
%  \begin{center}
%    \FigureFile(80mm,80mm){fig1.eps}
    %%% \FigureFile(width,height){filename}
%  \end{center}
%  \caption{This is the first figure.}\label{fig:sample}
%\end{figure}

%\setcounter{table}{0}

\section{Observations and data reduction}
High-resolution spectroscopic observations of V1280 Sco were performed using the 8.2-m Subaru telescope equipped with the High Dispersion Spectrograph (HDS; see \cite{noguchi02} for details) from 2009 to 2012. This period corresponds to the long plateau phase shown in Fig. 1, which includes photometric data obtained in 2012 at Osaka Kyoiku University. Table 1 provides a journal of the observations, in which the observing dates are listed in the universal time at the start time of the exposure. We have obtained 21 spectra under 5 configurations of the spectrograph, which cover the wavelength region from 3030 \AA~to 8070 \AA~ and provide the typical resolving power of $R$ $\cong$ 60\,000 at H$\alpha$. The slit width is 0".6 (0.3 mm). Data reduction was carried out using the IRAF\footnote[1]{IRAF is distributed by NOAO for Research in Astronomy, Inc., under cooperative agreement with the National Science Foundation.} software package in a standard manner. The wavelength calibration was performed using the Th-Ar comparison spectrum obtained during each observation. Spectrophotometric calibration was performed using the spectrum of HIP 83740 (A0 III) obtained at nearly same altitude of V1280 Sco on the same nights, with exceptions of the data obtained in 2009 May  and 2011 March due to hazy weather conditions. All spectra were converted to the helio-centric scale. Corrections for  interstellar extinction have not been applied.

\begin{figure}[hbt] %Figure 1
\begin{center}
 \FigureFile(85mm,85mm){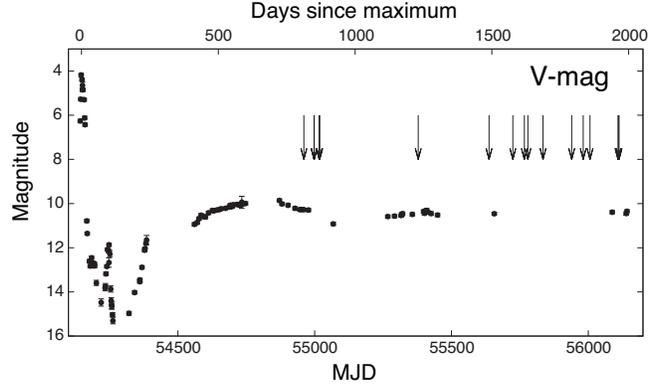}
\caption{Light curve of V1280 Sco observed at Osaka Kyoiku University. The epochs of our spectroscopic observations are indicated by arrows.}
\label{Fig.1}
\end{center}
\end{figure}

\begin{table}[h] %Table 1
\caption{Journal of spectroscopic observations of V1280 Sco.}
\begin{center}
\begin{tabular}{lrccc}
\hline
Date & UT & MJD & Exposure & Range \\
  &  h m& & sec & \AA\ \\
\hline
2009 & & & &\\
May 9 & 12 24 & 54\,960.517 & 900 & 4100-6860   \\
Jun. 15 & 12 36 & 54\,997.525 & 600 & 4100-6860    \\
Jun. 16 & 6 45 & 54\,998.282 & 900 & 3400-5110  \\
	& 11 48 & 54\,998.492 & 900 & 4110-6870   \\
	& 12 14 & 54\,998.510 & 900 & 5310-8070   \\
Jul. 4 & 10 51 & 55\,016.452 & 900 &  4100-6860  \\
Jul. 6 &  9 19 & 55\,018.389  & 1200 & 4100-6860  \\
	& 10 37 & 55\,018.443  & 600 & 3400-5110  \\

\hline
2010 & & & &\\
Jul. 1&  9 43  & 55\,378.405 & 1200 & 4130-6860 \\
\hline
2011  & & & &\\
Mar. 17 & 15 27 & 55\,637.644 & 930 &4100-6860  \\
Jun. 12 & 10 33 & 55\,724.440  & 1200 & 4100-6860 \\
Jul. 24 & 6 36 &  55\,766.275 & 1800 & 4100-6860  \\
Aug. 6  & 5 43 & 55\,779.239 & 1200& 4100-6860 \\
	  & 6 39 & 55\,779.277 & 1200 & 3030-4630 \\
Sep. 30  & 5 39 & 55\,834.236 & 2400 & 4130-6860 \\
\hline
2012  & & & &  \\
Jan. 13 & 16 17 & 55\,939.679 & 137 & 5080-7850\\
Feb. 24 & 15 29 & 55\,981.645 & 1800 &  4100-6860 \\
Mar. 20  & 14 41 & 56\,006.612 & 1200 & 4100-6860 \\
	& 15 14 & 56\,006.635 & 1800 & 3400-5110 \\
Jun. 30  & 7 53 & 56\,108.328 & 1200 & 3540-5250 \\
Jul. 4  & 10 3 & 56\,112.419 & 600 & 4100-6860 \\
\hline
\multicolumn{5}{l}{UT: Universal time at the start time of an exposure}\\
 \end{tabular} 
  \end{center}
\end{table}

\section{Confirmation of metastable He~{\sc i*} absorption lines}

We reported in Paper I the discovery of multiple high-velocity circumstellar absorption lines of Na~{\sc i} D2 and D1 on high-resolution spectra  observed in 2009. We analyzed additional spectra ranging from 3030 \AA~to 8070  \AA~ to look for similar absorption features associated with other elements. We found such blue-shifted multiple absorption lines corresponding to metastable He~{\sc i*} $\lambda$$\lambda$3188, 3889 and Ca~{\sc ii} H and K. Multiple absorption lines of Na~{\sc i} and Ca~{\sc ii} have been discussed in the literatures (e.g. \cite{williams08}; \cite{williams10}), those of the metastable He~{\sc i*} have never been reported in classical novae.

Figure 2 shows the spectral region between 3820 \AA~and 3900 \AA~ obtained on 2009 June 16, 2011 August 6, and 2012 March 20. These spectra show a continuum with H~{\sc i}, Fe~{\sc ii}, and Si~{\sc ii} lines in emission. While there are neither appearances nor disappearances of major emission lines, multiple absorption lines around 3880 \AA~ (shifted by $\sim$ $-$700 km s$^{-1}$ from 3889 \AA) are detected in 2011 and 2012. The feature is not observed in 2009. To examine whether blue-shifted components of H~{\sc i} $\lambda$3889 (H8) can be responsible for these absorption lines, we examine the presence of absorption features expected at the corresponding wavelengths for H~{\sc i} $\lambda$3835 (H9). There is no signature of absorption for the H9 line. Therefore, we guess the multiple absorption lines as those of the metastable He~{\sc i*} $\lambda$3889.

\begin{figure}[hbt] %Figure 2
\begin{center}
 \FigureFile(80mm,80mm){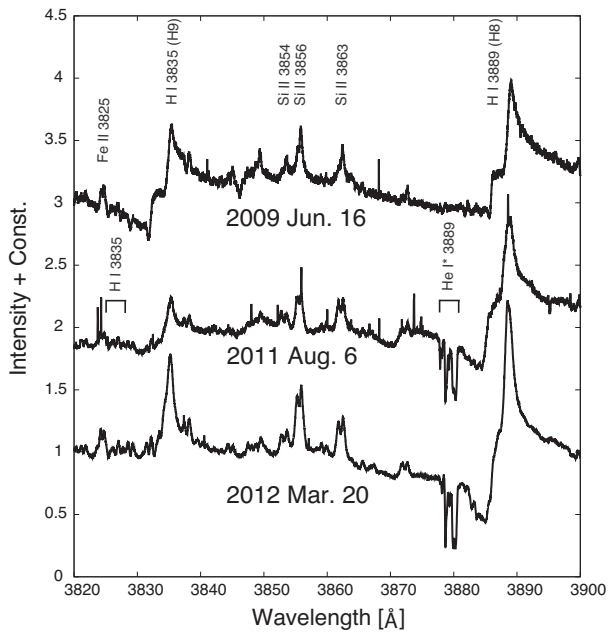}
\caption{Detection of multiple absorption lines around He~{\sc i*} $\lambda$3889 in 2011 and 2012. These absorption lines originate from the metastable He~{\sc I*}. Blue-shifted absorptions of the H~{\sc i} line at 3889 \AA~(H8) are negated because no corresponding absorptions  can be found at H9.}
\label{Fig.2}
\end{center}
\end{figure}

Since both He~{\sc i*} $\lambda$$\lambda$3188 and 3889 arise from the same metastable state 2~$^{3}$S, a similar set of absorption features is expected to be seen in He~{\sc i*} $\lambda$3188. Figure 3 shows a comparison of multiple high-velocity absorption lines in the He~{\sc i*} $\lambda$$\lambda$3188 and 3889 observed on 2011 August 6. While the signal-to-noise ratio for He~{\sc i*} $\lambda$3188 is rather low, we can see exact coincidences of four major absorption features (components A, B, L, and N). Thus, the identification of He~{\sc i} metastable lines in V1280 Sco can be definitely concluded. 

As far as we know, the He~{\sc i*} $\lambda$3889 absorption line has been reported only in the the recurrent nova U Sco \citep{johnston92}. However, the observed velocity in U Sco ($\sim$ $-$120 km s$^{-1}$) is significantly lower than those of V1280 Sco. \citet{johnston92} suggested that the He~{\sc i*}  absorption dose not originate in the ejected shells but from a corona around the secondary star in U Sco system. Hence, we conclude that our observation is the first detection of metastable He~{\sc i*} absorption lines originating in the ejected (circumstellar) gas around novae.
%At the moment, however, we can not determine whether the presence of  He~{\sc i*} absorption line components is unique to the V1280 Sco nova.

\begin{figure}[hbt] %Figure 3
\begin{center}
 \FigureFile(80mm,80mm){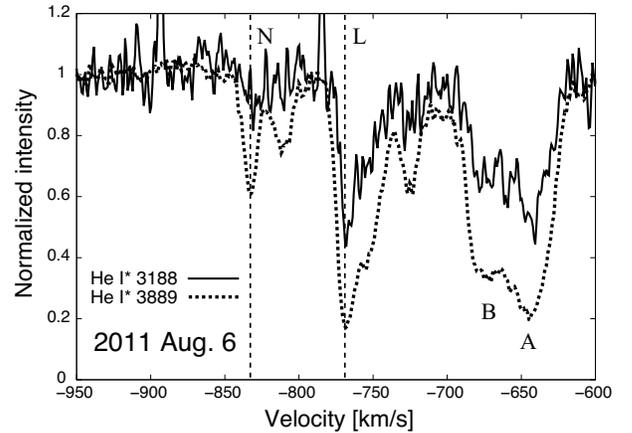}
\caption{Comparison of multiple blue-shifted high-velocity absorption components originating from metastable He~{\sc i*} $\lambda$3188 and He~{\sc i*} $\lambda$3889 observed on 2011 August 6. Both lines arise from the same triplet metastable level (2~$^{3}$S) of He~{\sc i}. Although the signal-to-noise level for He~{\sc i*} $\lambda$3188 is much lower, there is little doubt that both lines show the same absorbing components.} 
\label{Fig.3}
\end{center}
\end{figure}

\begin{figure}[hbt] %Figure 4
\begin{center}
 \FigureFile(80mm,80mm){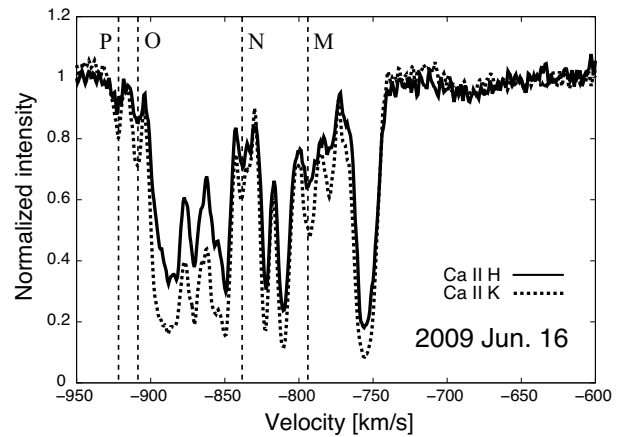}
\caption{Comparison of multiple blue-shifted high-velocity absorption components originating from Ca~{\sc ii} H and K observed on 2009 June 16.} 
\label{Fig.4}
\end{center}
\end{figure}

\section{Co-existence of He~{\sc i*}, Na~{\sc i}, and Ca~{\sc ii} absorption lines}
We have discovered at least eleven discrete high-velocity absorption components in the Na~{\sc i} D doublet on the high-resolution spectra observed in 2009 (in Paper I). We find that similar high-velocity absorption lines can be seen in Ca~{\sc ii} H and K on the high-resolution spectra observed in 2009 (Fig. 4). It is interesting to find at least four new absorption features in Ca~{\sc ii} H and K lines which have not been noticed in Paper I. They are components M, N, O, and P in Fig. 4. The component P is displaced by $-$922 km s$^{-1}$, which is higher by $\sim$ 40 km s$^{-1}$ than the fastest component found in the Na~{\sc i} D absorptions.

\begin{figure}[hbt] %Figure 5
\begin{center}
 \FigureFile(80mm,80mm){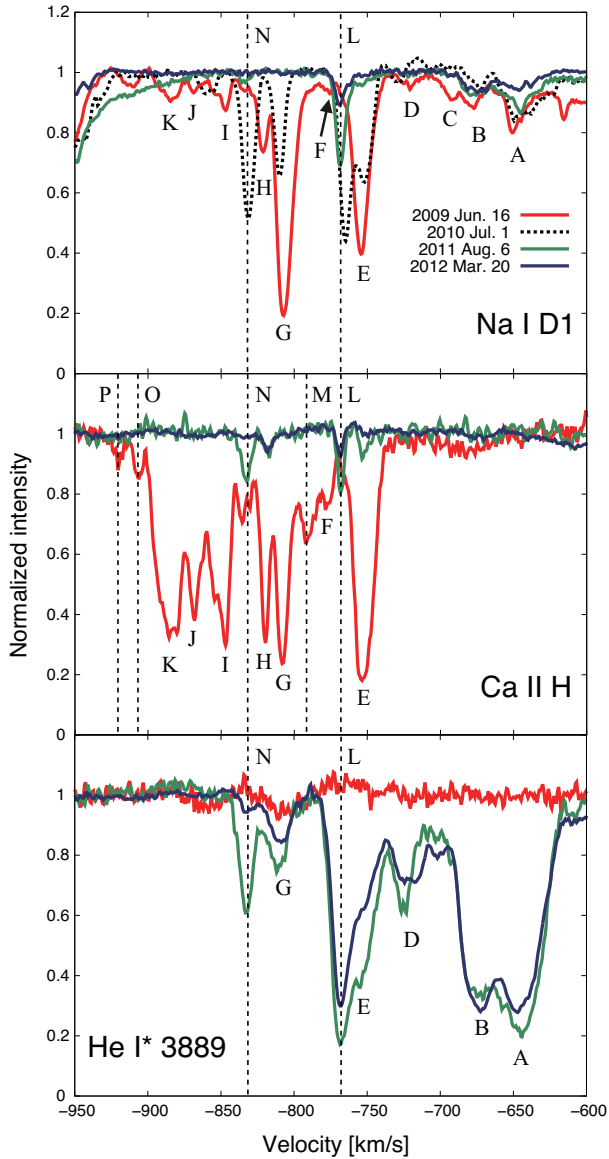}
\caption{Developments of absorption profiles of Na~{\sc i} D1, Ca~{\sc ii} H, and He~{\sc i*} $\lambda$3889 from 2009 to 2012. Components A, B, C, D, E, F, G, H, I, J, and K are introduced which are associated with those in Paper I.  Components L, M, N, O, and P are the same as Figs. 3 and 4. We have no data for Ca~{\sc ii} H and He~{\sc i*} $\lambda$3889 in 2010.} 
\label{Fig.5}
\end{center}
\end{figure}

\begin{figure}[hbt] %Figure 6
\begin{center}
 \FigureFile(80mm,80mm){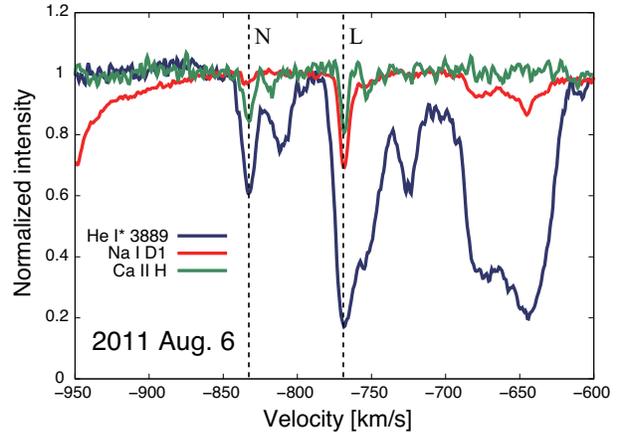}
\caption{Profiles of blue-shifted high-velocity absorption components of He~{\sc i*} $\lambda$3889, Na~{\sc i} D1, and Ca~{\sc ii} H on 2011 August 6. Each absorption component is noted as in Fig. 3. Components L and N are seen in all the species on 2011 August 6.} 
\label{Fig.6}
\end{center}
\end{figure}

Figure 5 shows comparison of absorption profiles of Na~{\sc i} D1, Ca~{\sc ii} H, and He~{\sc i*} $\lambda$3889 observed from 2009 through 2012. Components A through K are the same as noted in Paper I. For He~{\sc i*} lines, four major absorption components are identified in the 2011 spectra (noted as A, B, L, and N in Fig. 3). Two new absorption components L and N of Na~{\sc i} D developed in 2010. Temporal variations in both wavelengths and strengths among these spectra illustrate complex changes in column density and velocity for each component. Measurements of the velocities and the depths of discrete absorption components are tabulated in Table 2. Errors in measurements are estimated to be $\pm$ 5 km s$^{-1}$ in velocity and $\pm$ 0.02 in depth.

On the component G of Na~{\sc i} D1, we find that the depth becomes shallower (from 0.81 to 0.34), and the line center is shifted (from −806.1 km s$^{-1}$ to −809.2 km s$^{-1}$) from 2009 to 2010. Similar changes are noted in the components A, B, E, and L. These findings might indicate that the broader components consist of several narrow lines coming from smaller clumpy clouds having different velocities, and the observed variation in velocity can be interpreted as due to change of contributing ratio of each clumpy component. There seems no acceleration mechanism for the particular components during the early period as suggested by \citet{williams11}. The fact that the deepest absorption lines for all elements show a certain level of intensity ($\sim$ 0.2) at line centers indicates that the covering factor of combined components is at least larger than 0.8 (see Sect. 5.2).

Metastable He~{\sc i*} absorption lines are observed in other astronomical objects such as Seyfert galaxies (e.g. Markarian 231: \cite{boksenberg77}; NGC 4151: \cite{anderson69}; \cite{anderson74}), quasars (e.g. QSO 2359$-$1241: \cite{arav01}), or nebula (e.g. the Orion Nebula: \cite{wilson59}).
%These objects have regions of low-density interstellar or circumstellar matter.
\citet{anderson74} gives a conclusion that He~{\sc i*} absorption features arise in the region characterized by higher electron density ( $>$ 10$^{6}$ cm$^{-3}$) in NGC 4151.
The Seyfert galaxy Markarian 231 shows multiple absorption lines of He~{\sc i*} $\lambda$3889, Na~{\sc i} D and Ca~{\sc ii} H and K \citep{boksenberg77} and similar features -- co-existence of He~{\sc i*}, Na~{\sc i}, and Ca~{\sc ii} originating in absorbing shells -- are observed in components L and N of V1280 Sco (Fig. 6). 
%suggests that V1280 Sco also has circumstellar matter illuminated by a dilute ionizing radiation field from a high-temperature source.
Figure 6 shows that He~{\sc i*} absorptions are much stronger than either Na~{\sc i} or Ca~{\sc ii} absorptions in all of the absorption systems, which demonstrates that the most parts of the absorbing clouds are strongly ionized by a high-temperature radiation source.

\section{Discussion}
Recent high-resolution spectroscopic observations have revealed that narrow, time-variable Na~{\sc i} D absorption features are exhibited in some kinds of cataclysmic variable systems, i.e., novae (e.g. V1280 Sco: Paper I; RS Oph: \cite{patat11}; T Pyx: \cite{shore11}; V407 Cyg: \cite{shore12}) and type Ia supernovae (e.g. SN 2006X: \cite{patat07}; SN 1999cl: \cite{blondin09}; SN 2007le: \cite{simon09};  PTF 11kx:  \cite{dilday12}). The origins of the absorption lines are presently controversial, however, these lines could be used to investigate the structure and evolution of circumstellar environment around progenitor systems with the implication for physical connections among various types of cataclysmic variables.

According to \citet{williams10}, multiple high-velocity absorption lines of Na~{\sc i} D doublet are observed in $\sim$ 80 \% of classical novae. These lines weaken and disappear  within several weeks/months after the maximum light. Most of the novae with such absorption lines exhibit a gradual outward acceleration in the weeks following the maximum light. While being common among novae, the physical nature of these lines have been scarcely discussed. Two scenarios for the origins of these absorption lines are proposed in \citet{williams11}: (1) a pre-existing circumbinary reservoir before the nova outburst, or (2) outflowing high-density blobs associated with post-outburst ejecta. Since multiple absorption systems are expected to appear and disappear in different ways, more time-sequence observational data of many novae could discriminate between the two scenarios \citep{williams11}.
%However, the two scenarios have some problems to be solved: (1) the required mass of pre-existing circumbinary gas to satisfy the observed line intensity is significantly large -- an order of 10$^{-5}$ M$_{\odot}$; however, there seems no way to explain this large amount which is comparable to typical ejected mass of nova outburst, (2) the number of absorption components originate from post-outburst ejecta is expected to increase when the photosphere decreases its size in the late stage; however, observations show that the number of multiple absorption lines reduces with time.
Another interpretation accounting for the observed behavior of absorption lines is presented by \citet{shore11}. They argue that the time sequence of absorption lines can be explained in a situation where recombination and/or ionization fronts move outward in the expanding gas. The scenario is based on the study of recurrent nova T Pyx, which exhibits massive ejecta in each event.

In this section, we discuss the origins of discrete absorption lines and structure of ejected shell of V1280 Sco using metastable He~{\sc i*} absorption lines along with Na~{\sc i} D doublet and Ca~{\sc ii} H and K. This new approach can be widely employed to probe the circumstellar structure of cataclysmic variables.

\subsection{Survival time of discrete absorption lines}
Some components corresponding to Na~{\sc i} D1 and Ca~{\sc ii} H (components A and B) weakened significantly from 2009 to 2012. On the other hand, these two components of He~{\sc i*} $\lambda$3889 were absent in 2009 and remarkably grown in 2011 and 2012. Survival time of these absorption lines in V1280 Sco is an order of years, which is much longer than those observed among fast novae (an order of weeks or months). Furthermore, He~{\sc i*} $\lambda$3889 absorption lines were detected for the first time in 2011, four years after the maximum light. These behavior associated with He~{\sc i*} can be understood that the number of ultraviolet photons had increased significantly to produce singly-ionized helium as the central photosphere shrinks to become hotter. Consequently, we postulate that the complex evolutions of multiple absorption lines are due to combined changes in physical conditions, such as the density, recombination and ionization rate, for each component. Considering that V1280 Sco had taken a very long time to emit ultraviolet radiation extensively \citep{naito12}, the most significant parameter extending the survival time of multi absorption lines of Na~{\sc i} D and Ca~{\sc ii} H and K -- low excitation lines-- could be the ionization rate caused by a change in the temperature of the photosphere.

\begin{table*} %Table 2
\caption{Velocities and depths of absorption components.}
\begin{center}
\begin{tabular}{llcccccccc}
\hline
\multicolumn{2}{c}{Component$\dagger$} & \multicolumn{4}{c}{Velocity} 	& \multicolumn{4}{c}{Depth} \\
	&	&  \multicolumn{4}{c}{(km s$^{-1}$)} & 	 \\
	&				&  2009 		& 2010 		& 2011		& 2012 		& 2009 	& 2010 	& 2011 	& 2012\\
\hline
A 	& Na~{\sc i} D1	 & 	$-650.0$		& $-649.0$	& $-644.9$	& $-646.9$	& $0.20$	& $0.16$	& $0.13$ & $0.06$ \\
	& Ca~{\sc ii} H	 & 	$\dots$		&  $	$		&  $\dots$		& $\dots$		& $\dots$	& $	$	& $\dots$ & $\dots$ \\
 	& He~{\sc i*} 3889	 & $\dots$		&  $	$		&  $-644.4$	& $-647.2$	& $\dots$	& $	$	& $0.80$ & $0.72$ \\
B	& Na~{\sc i} D1	 & 	$-678.5$		&  $-671.9$	&  $-676.0$	& $-673.3$	& $0.12$	&  $0.06$	& $0.07$ & $0.06$ \\
	& Ca~{\sc ii} H	 & 	$\dots$		&  $	$		&  $\dots$		& $\dots$		& $\dots$	&  $	$	& $\dots$ & $\dots$ \\
 	& He~{\sc i*} 3889	& $\dots$		&  $	$		&  $-672.2$	& $-672.4$	& $\dots$	&  $	$	& $0.67$ & $0.72$ \\
C 	& Na~{\sc i} D1	 & 	$-691.3$		&  $\dots$		&  $\dots$		& $\dots$		&  $0.09$	& $\dots$	& $\dots$	& $\dots$ \\
	& Ca~{\sc ii} H	 & 	$\dots$		&  $	$		&  $\dots$		& $\dots$		& $\dots$& $	$	& $\dots$ & $\dots$ \\
 	& He~{\sc i*} 3889	 & $\dots$		&  $	$		&  $\dots$		& $\dots$		& $\dots$	& $	$	& $\dots$ & $\dots$ \\
D 	& Na~{\sc i} D1	 & 	$-720.2$		&  $\dots$		&  $\dots$		& $\dots$		&  $0.06$	& $\dots$	& $\dots$	& $\dots$ \\
	& Ca~{\sc ii} H	 & 	$\dots$		&  $	$		&  $\dots$		& $\dots$		& $\dots$& $	$	& $\dots$ & $\dots$ \\
 	& He~{\sc i*} 3889	 & $\dots$		&  $	$		&  $-724.6$	& $-720.8$	& $\dots$	& $	$	& $0.39$ & $0.29$ \\
E 	& Na~{\sc i} D1	 & 	$-753.8$		&  $-751.7$	&  $\dots$		& $\dots$		&  $0.61$	& $0.37$	& $\dots$	& $\dots$ \\
	& Ca~{\sc ii} H	 & 	$-753.7$		&  $	$		&  $-753.1$	& $\dots$		& $0.82$	&$	$	& $0.08$ & $\dots$ \\
 	& He~{\sc i*} 3889	 & $\dots$		&  $	$		&  $-754.7$	& $\dots$		& $\dots$	& $	$	& $0.64$ & $\dots$ \\
F 	& Na~{\sc i} D1	 & 	$-776.7$		&  $\dots$		&  $\dots$		& $\dots$		&  $0.07$	& $\dots$	& $\dots$	& $\dots$ \\
	& Ca~{\sc ii} H	 & 	$-779.6$		&  $	$		&  $\dots$		& $\dots$		& $0.24$& $	$	& $\dots$ & $\dots$ \\
 	& He~{\sc i*} 3889	 & $\dots$		&  $	$		&  $\dots$		& $\dots$		& $\dots$	& $	$	&$\dots$ & $\dots$ \\
G 	& Na~{\sc i} D1	 & 	$-806.1$		&  $-809.2$	&  $\dots$		& $\dots$		& $0.81$	& $0.34$	& $\dots$	& $\dots$ \\
	& Ca~{\sc ii} H	 & 	$-807.5$		&  $	$		&  $\dots$		& $\dots$		& $0.76$	& $	$	& $\dots$ & $\dots$ \\
 	& He~{\sc i*} 3889	 & $\dots$		&  $	$		&  $-810.2$	& $-808.3$	& $\dots$	& $	$	& $0.25$ & $0.16$ \\
H 	& Na~{\sc i} D1	 & 	$-820.9$		&  $\dots$		&  $\dots$		& $\dots$		& $0.27$	&  $\dots$ &$\dots$	& $\dots$ \\
	& Ca~{\sc ii} H	 & 	$-818.8$		&  $	$		&  $-817.5$	& $-817.3$	& $0.69$	& $	$	& $0.06$ & $0.06$ \\
 	& He~{\sc i*} 3889	 & $\dots$		&  $	$		&  $\dots$		& $\dots$		& $\dots$	& $	$	& $\dots$ & $\dots$ \\
I	& Na~{\sc i} D1	 & 	$-846.3$		&  $\dots$		&  $\dots$		& $\dots$		&$0.13$	&  $\dots$	& $\dots$ 	& $\dots$ \\
	& Ca~{\sc ii} H	 & 	$-846.8$		&  $	$		&  $\dots$		& $\dots$		& $0.70$	&$	$	& $\dots$ & $\dots$ \\
 	& He~{\sc i*} 3889	 & $\dots$		&  $	$		&  $\dots$		& $\dots$		& $\dots$	& $	$	& $\dots$ & $\dots$ \\
J 	& Na~{\sc i} D1	 & 	$-868.2$		&  $\dots$		&  $\dots$		& $\dots$		& $0.07$	&  $\dots$	& $\dots$	& $\dots$ \\
	& Ca~{\sc ii} H	 & 	$-867.9$		&  $	$		&  $\dots$		& $\dots$		& $0.62$	& $	$	& $\dots$ & $\dots$ \\
 	& He~{\sc i*} 3889	 & $\dots$		&  $	$		&  $\dots$		& $\dots$		& $\dots$	& $	$	& $\dots$& $\dots$ \\
K 	& Na~{\sc i} D1	 & 	$-882.9$		& $\dots$		&  $\dots$		& $\dots$		& $0.09$	&  $\dots$	& $\dots$ 	& $\dots$ \\
	& Ca~{\sc ii} H	 & 	$-883.2$		&  $	$		&  $\dots$		& $\dots$		& $0.67$	& $	$	& $\dots$ & $\dots$ \\
 	& He~{\sc i*} 3889	 &  $\dots$	&  $	$		& $\dots$		& $\dots$		& $\dots$	& $	$ 	& $\dots$ & $\dots$\\
\hline	
L 	& Na~{\sc i} D1	 & 	$\dots$		&  $-763.9$	&  $-767.5$	& $-768.5$	& $\dots$	&  $0.56$	& $0.30$ & $0.11$ \\
	& Ca~{\sc ii} H	 & 	$\dots$		&  $	$		&  $-767.4$	& $-768.2$	& $\dots$	&  $	$	& $0.19$ & $0.07$ \\
 	& He~{\sc i*} 3889	&$\dots$		&  $	$		&  $-767.0$	& $-767.5$	& $\dots$	&  $	$	& $0.83$ & $0.70$ \\
M 	& Na~{\sc i} D1	 & 	$\dots$		& $\dots$		&  $\dots$		& $\dots$		& $\dots$	&  $\dots$	& $\dots$ 	& $\dots$ \\
	& Ca~{\sc ii} H	 & 	$-790.9$		&  $	$		&  $\dots$		& $\dots$		& $0.35$	& $	$	& $\dots$ & $\dots$ \\
 	& He~{\sc i*} 3889	 &  $\dots$	&  $	$		& $\dots$		& $\dots$		& $\dots$	& $	$ 	& $\dots$ & $\dots$\\	
N 	& Na~{\sc i} D1	 & 	$-833.6$		&  $-830.5$	&  $-833.1$	& $\dots$		& $0.06$	&  $0.46$	& $0.03$ & $\dots$ \\
	& Ca~{\sc ii} H	 & 	$-833.7$		&  $	$		&  $-832.4$	& $\dots$		& $0.27$	&  $	$	& $0.16$ & $\dots$ \\
 	& He~{\sc i*} 3889	 &$\dots$		&  $	$		&  $-832.5$	& $-832.7$	& $\dots$	&  $	$	& $0.39$ & $0.06$ \\	
O 	& Na~{\sc i} D1	 & 	$\dots$		& $\dots$		&  $\dots$		& $\dots$		& $\dots$	&  $\dots$	& $\dots$ 	& $\dots$ \\
	& Ca~{\sc ii} H	 & 	$-905.8$		&  $	$		&  $\dots$		& $\dots$		& $0.13$	& $	$	& $\dots$ & $\dots$ \\
 	& He~{\sc i*} 3889	 &  $\dots$	&  $	$		& $\dots$		& $\dots$		& $\dots$	& $	$ 	& $\dots$ & $\dots$\\	
P 	& Na~{\sc i} D1	 & 	$\dots$		& $\dots$		&  $\dots$		& $\dots$		& $\dots$	&  $\dots$	& $\dots$ 	& $\dots$ \\
	& Ca~{\sc ii} H	 & 	$-922.4$		&  $	$		&  $\dots$		& $\dots$		& $0.11$	& $	$	& $\dots$ & $\dots$ \\
 	& He~{\sc i*} 3889	 &  $\dots$	&  $	$		& $\dots$		& $\dots$		& $\dots$	& $	$ 	& $\dots$ & $\dots$\\	
\hline
\multicolumn{10}{l}{$\dagger$: Components A, B, C, D, E, F, G, H, I, J, and K are labeled as in Paper I.}\\
\multicolumn{10}{l}{ }\\
 \end{tabular} 
  \end{center}
\end{table*}

\begin{table*} %Table 3
\caption{Measured parameters for components of A, B, and L using both metastable He~{\sc i*} $\lambda$$\lambda$3188 and 3889 lines.}
\begin{center}
\begin{tabular}{ccccccccc}
\hline
Component & Velocity 	&  $C_\mathrm{f}$ &	$\tau_{3188}$	& $\tau_{3889}$ &	$b$ 				& $N(2~^3\mathrm{S})$ 	& \multicolumn{2}{c}{$N(\mathrm{He}^{+})$} \\
		& (km s$^{-1}$) & 				&			& 			 & ($10^{6}$ cm s$^{-1}$)	&  ($10^{13}$ cm$^{-2}$)	&  \multicolumn{2}{c}{($10^{19}$ cm$^{-2}$)} \\
		& 			& 				&			& 			 & 					&  					& $n_{\mathrm e}$ = 10$^{5}$ & $n_{\mathrm e}$ = 10$^{6}$\\
\hline
A 		 & $-644.4$	& 0.93			& 	0.98			& 2.94			& 	1.2	$\pm$ 0.5		& 9.3 $\pm$ 3.9 	& 1.7 & 1.6\\
B		 &  $-672.2$	& 0.95			& 	0.63			& 1.90			& 	1.6	$\pm$ 0.5		& 7.8 $\pm$ 2.4 	& 1.4 & 1.3\\
L 		 & $-767.0$	& 0.86			&	1.19			 & 3.56			& 	1.0    $\pm$ 0.6		& 9.8 $\pm$ 5.9 	& 1.7 & 1.7 \\
\hline
%\multicolumn{4}{l}{$\ast$: Spectra are shown in the online material}\\
 \end{tabular} 
  \end{center}
\end{table*}

\subsection{Column density of He (2~$^{3}$S)}
Metastable He~{\sc i*} lines are often used to derive physical parameters of nebulosity because processes of transition are well understood. Of the recombinations to excited levels of He~{\sc i}, approximately three-quarters are led to the triplet levels and approximately one-quarters are led to the singlet levels. All electrons in the triplets are led ultimately to 2~$^{3}$S through downward radiative transitions, which is highly metastable \citep{osterbrock06}. Since both He~{\sc i*} $\lambda$3188 and He~{\sc i*} $\lambda$3889 are transitions from the same metastable level, the ratio of optical depths is fixed by atomic physics, provided that the lines are not saturated.
%We can use these ratios to solve for the parameters for each absorption component.
If we adopt the partial covering model used in \citet{leighly11}, the following relations are obtained:
\begin{eqnarray}
R_{3188} = (1 - C_{\rm f}) + C_{\rm f}  e^{-\tau_{3188}}
\end{eqnarray}
\begin{eqnarray}
R_{3889} = (1 - C_{\rm f}) + C_{\rm f}  e^{-\tau_{3889}}
\end{eqnarray}
where $R$ is the ratio of the observed spectrum intensity to the continuum, $C_{\mathrm f}$ is the fraction of  the continuum source which is covered by the absorbing clumpy gas  (the covering factor), and $\tau_{3188}$ and $\tau_{3889}$ are the optical depths of He~{\sc i*} $\lambda$3188 and He~{\sc i*} $\lambda$3889, respectively. All of these parameters are functions of velocity. When we know the optical depth $\tau$ at line center, the column density $N$ can be derived from the following expression given in \citet{rudy85}:
\begin{eqnarray}
N = (\tau b)/(0.015 \lambda f_{12})
\end{eqnarray}
where $\lambda$ is the wavelength in cm, $f_{12}$ is the oscillator strength, and $b$ is a measure of the velocity width in units of cm s$^{-1}$.
The column density of the 2~$^{3}$S level can be transformed into the column density of He$^{+}$ using the relation given in \citet{clegg87}:
\begin{eqnarray}
\frac{N(2~^3\mathrm{S})}{N(\mathrm{He}^{+})}= \frac{5.79 \times 10^{-6} t_{\mathrm e}^{-1.18}}{1+3110 t_{\mathrm e}^{-0.51} n_{\mathrm e}^{-1}}
\end{eqnarray}
where $t_{\mathrm e}$ is the electron temperature in 10$^4$ K and $n_{\mathrm e}$ is the electron density in cm$^{-3}$. When we use a typical value of 10$^4$ K for  the electron temperature, the ratio of {$N$(2~$^{3}$S) to ${N(\mathrm{He}^{+})}$ can be confined $\sim$ 10$^{-6}$ for the low electron density ($n_{\mathrm e}$ $\sim$ 10$^5$ $-$ 10$^6$; see Appendix 1 for details). Measured parameters for components of A, B, and L are listed in Table 3, where we calculate for two cases of $n_{\mathrm e}$ = 1 $\times$10$^5$ and $n_{\mathrm e}$ = 1 $\times$ 10$^6$, respectively. We find that the column density of He$^{+}$ of each component is $\sim$ 10$^{19}$ cm$^{-2}$ for both electron densities.

\subsection{Structure and mass of ejected shell}
 \citet{chesneau12} obtained high-spatial resolution mid-infrared images with the Very Large Telescope from 2009 to 2011, showing the dusty nebular around V1280 Sco is bipolar. They suggested that the intensities of two bipolar lobes are different due to an absorption effect; the southern lobe is closer to the observer than the northern one. It is implied that a large part of ejected mass is included in these bipolar lobes. Combined with the presence of absorbing gases on the line-of-sight shown in our observations, we can describe the structure of ejected shell as Fig. 7. We aim to estimate the mass of ejected shell assuming that the components showing He~{\sc i*} $\lambda$$\lambda$3188 and 3889 originate from clumpy gas ejected at the time of the explosion in 2007. As noted in Sect. 5.2, each absorption component covers most of the continuum radiating region (photosphere and free-free emission region) on the line-of-sight and extended $\sim$ 1 $\times$ 10$^{9}$ km ($\sim$ 10 AU) in the past $\sim$1600 days with the velocity dispersion (FWHM) of $\Delta$10 km s$^{-1}$. To achieve the column density of He$^{+}$ of $\sim$ 10$^{19}$ cm$^{-2}$, the following physical condition is needed for the absorbing clumpy gas; (i) $n_{\mathrm e}$ $\sim$ 10$^6$ for the solar abundance of He or (ii) $n_{\mathrm e}$ $\sim$ 10$^5$ for helium-rich composition. If condition (ii) is the case, the clumpy gas could consist of helium-rich component ejected after hydrogen burning, which is consistent with normal nova phenomena. Supposing the density in individual clumpy gas is uniform, we can estimate the total mass of ejected shell to be on the order of 10$^{-4} $M$_{\odot}$ by integrating absorption components ranging from $- 650$ km s$^{-1}$ to $- 920$ km s$^{-1}$. This is derived by a distance-independent method and is consistent with the results of \citet{chesneau08} and \citet{naito12}.

\begin{figure}[hbt] %Figure 7
\begin{center}
 \FigureFile(90mm,90mm){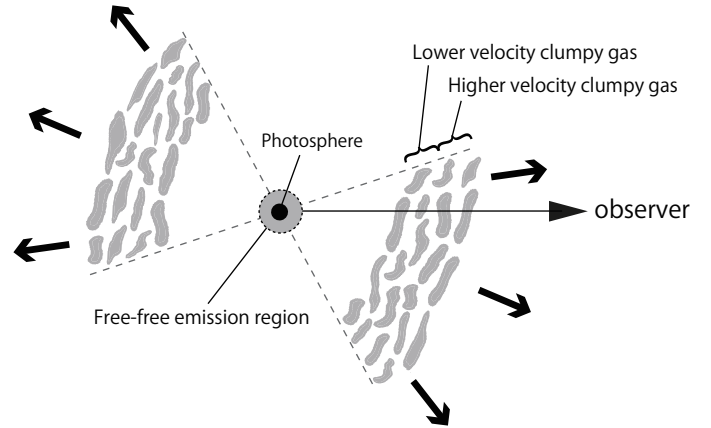}
\caption{Schematic of ejected shell (clumpy gas) producing absorption lines of metastable He~{\sc i*} $\lambda$$\lambda$3188 and 3889. Components A  and B originate in lower velocity clumpy gases and components L and N originate in higher velocity clumpy gases. Thin broken lines delineate boundaries of the nebula (\cite{chesneau12}).} 
\label{Fig.7}
\end{center}
\end{figure}

\section{Conclusions}
We performed high-resolution spectroscopic observations of V1280 Sco from 2009 to 2012 and discovered multiple absorption lines corresponding to metastable He~{\sc i*} $\lambda$$\lambda$3188 and 3889. These lines have corresponding absorption components in blue-shifted Na~{\sc i} D doublet and Ca~{\sc ii} H and K on the spectra of V1280 Sco. Multiple absorption lines originate in ejected shells consisting of clumpy components which cover a large part of the continuum radiating region (photosphere and free-free emission region) on the line-of-sight.
The complex evolution of multiple absorption lines can be explained by the result of changing in density and  recombination/ionization rate when the shell expands and the photosphere shrinks to become hotter.
We estimate the total mass of ejected shell to be on the order of $\sim$ 10$^{-4}$ M$_{\odot}$ using metastable He~{\sc i*} $\lambda$$\lambda$3188 and 3889 absorption lines.

%%%%%%%%%%%%%%%%%%%%%%%%%%%%%%%%%%%%%%%

%\begin{longtable}{lll}
%  \caption{Sample of ``longtable"}\label{tab:LTsample}
%  \hline              
%  name & value1 & value2 \\ 
%\endfirsthead
%  \hline
%  name & value & value2  \\
%\endhead
%  \hline
%\endfoot
%  \hline
%\endlastfoot
%  \hline
%  aaaaa & bbbbb & ccccc \\
%  ...... & ..... & ..... \\
%  ...... & ..... & ..... \\
%  ...... & ..... & ..... \\
%  xxxxx & yyyyy & zzzzz \\
%\end{longtable}

\bigskip
We are grateful to students at Osaka Kyoiku University for performing the photometric observations in 2011 and 2012.
We sincerely thank Takashi Iijima and Hitoshi Yamaoka for useful comments.
Thanks are also due to Kazunori Ishibashi and Shinjirou Kouzuma for careful reading of the manuscript and suggestions.
This work was supported by the Global COE Program of Nagoya University "Quest for Fundamental Principles in the Universe (QFPU)" from JSPS and MEXT of Japan.

\appendix
%\section{Method of .....}
%\section{Approximation of ...}
%\section*{Complete data}
\section{Estimation of the electron density in absorption components}
According to \citet{williams11}, velocities of multiple absorption lines almost always correspond to half the velocities measured from forbidden emission line widths. This fact supports an assumption that these absorption and forbidden lines arise from the same region. Intensity ratios of forbidden lines for [O~{\sc ii}] and [O~{\sc iii}]  are expressed as functions of the electron density ($n_{\rm e}$) and electron temperature ($t_{\rm e}$) as below. \citet{peimbert67} gives the expression for  [O~{\sc ii}];
\begin{eqnarray}
\frac{I(\lambda 7320 + \lambda7330)}{I(\lambda 3726 + \lambda 3729)} = \frac{\epsilon}{13.70} \frac{1+23.8x(1+0.23\epsilon)+61.2x^{2}(1+0.61\epsilon+0.07\epsilon^{2})}{1+0.13\epsilon+5.3x(1+0.60\epsilon+0.07\epsilon^{2})},
\end{eqnarray}
where
\begin{eqnarray}
x=10^{-2}n_{\rm e}{t_{\rm e}}^{-1/2},
\end{eqnarray}
and
\begin{eqnarray}
\epsilon = \exp({-19600/t_{\rm e}}).
\end{eqnarray}
\citet{osterbrock06} give the expression for [O~{\sc iii}];
\begin{eqnarray}
\frac{I(\lambda 4363)}{I(\lambda 4959 + \lambda 5007)} = \frac{1+4.5\times10^{-4}n_{\rm e}/{t_{\rm e}}^{1/2}}{7.90\exp(3.29\times10^{4}/t_{\rm e})}.
\end{eqnarray}
When the intensity ratios of  [O~{\sc ii}] and  [O~{\sc iii}] are measured, the electron density and the electron temperature in absorption components can be derived.

Hence [O~{\sc ii}] 3726, 3729 and [O~{\sc iii}] 4363 lines are not detected in our spectra, the intensity ratios are not able to be measured exactly. However, we can set a limitation on the physical parameters by considering the detection limits for such weak forbidden lines. When an interstellar correction of $B(E-V) = 0.4$ is applied \citep{naito12}, the condition of both ${I(\lambda 7320 + \lambda7330)}/{I(\lambda 3726 + \lambda 3729)} > 0.86$  and ${I(\lambda 4363)}/{I(\lambda 4959 + \lambda 5007)} < 0.015$ is set to be satisfied. Adopting $\sim$ $10^{4}$ K as a typical value for the electron temperature, the electron density can be confined somewhere between $10^{5}$ and $10^{6}$ cm$^{-3}$ (see Fig. 8).

\begin{figure}[hbt] %Figure 8
\begin{center}
 \FigureFile(80mm,80mm){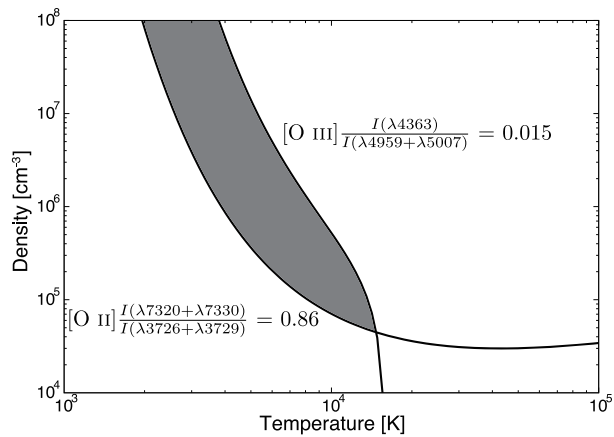}
\caption{Permitted area for the electron temperature and the electron density. The colored zone between two lines satisfies the conditions from [O II] and [O III] forbidden lines.} 
\label{Fig.8}
\end{center}
\end{figure}

%%%
% See the manual for the detail.
%%%


\begin{thebibliography}{}
% Journals(e.g. A\&A,ApJ,AJ,NMRAS,PASP ...)
% Authors, Year, Journal, Vol#, Page#
% Journal Title Abbreviation >> http://www.asj.or.jp/pasj/Jabb.html

\bibitem[Anderson \& Kraft(1969)]{anderson69}
{Anderson},~K.~S., \& {Kraft},~R.~P. 1969, \apj, 158, 859

\bibitem[Anderson(1974)]{anderson74}
{Anderson},~K.~S. 1974, \apj, 189, 195

\bibitem[Arav et al.(2001)]{arav01}
{Arav},~N., {Brotherton},~M.~S., {Becker},~R.~H., {Gregg},~M.~D., {White},~R.~L., {Price},~T., \& {Hack}, W. 2001, \apj, 546, 140

\bibitem[Blondin et al.(2009)]{blondin09}
{Blondin},~S., {Prieto},~J.~L., {Patat},~F., {Challis},~P., {Hicken},~M., {Kirshner},~R.~P., {Matheson},~T., \& {Modjaz},~M. 2009, \apj, 693, 207

\bibitem[Boksenberg et al.(1977)]{boksenberg77}
{Boksenberg},~A., {Carswell},~R.~F., {Allen},~D.~A., {Fosbury},~R.~A.~E., {Penston},~M.~V., \& {Sargent},~W.~L.~W. 1977, \mnras, 178, 451

\bibitem[Chesneau et al.(2008)]{chesneau08}
{Chesneau}, O., \etal\ 2008, \aap, 487, 223

\bibitem[Chesneau et al.(2012)]{chesneau12}
{Chesneau}, O., \etal\ 2012, \aap, 545, A63

\bibitem[Clegg(1987)]{clegg87}
{Clegg}, R.~E.~S. 1987, \mnras, 229, 31p

\bibitem[Dilday et al.(2012)]{dilday12}
{Dilday},~B., \etal\ 2012, Science, 337, 942

\bibitem[Hachisu \& Kato(2006)]{hachisu06}
{Hachisu},~I., \& {Kato},~M. 2006, \apjs, 167, 59

\bibitem[Johnston \& Kulkarni(1992)]{johnston92}
{Johnston},~H.~M., \& {Kulkarni},~S.~R. 1992, \apj, 396, 267

\bibitem[Kato \& Hachisu(1994)]{kato94}
{Kato}, M., \& {Hachisu}, I. 1994, \apj, 437, 802

\bibitem[Leighly et al.(2011)]{leighly11}
{Leighly},~K.~M., {Dietrich},~M., \& {Barber},~S. 2011, \apj, 728, 94

\bibitem[Naito et al.(2012)]{naito12}
{Naito},~H., \etal\ 2012, \aap, 543, A86

\bibitem[Noguchi et al.(2002)]{noguchi02}
{Noguchi},~K., \etal\ 2002, \pasj, 54, 855

\bibitem[Osterbrock \& Ferland(2006)]{osterbrock06}
{Osterbrock},~D.~E., \& {Ferland},~G.~J. 2006, in Astrophysics of gaseous nebulae and active galactic nuclei, 2nd.,~ed.~D.E.~Osterbrock \& G.J.~Ferland (Sausalito, California: University Science Books)

\bibitem[Patat et al.(2007)]{patat07}
{Patat},~F., \etal\ 2007, Science, 317, 924

\bibitem[Patat et al.(2011)]{patat11}
{Patat},~F., {Chugai},~N.~N., {Podsiadlowski},~P., {Mason},~E., {Melo},~C., {Pasquini},~L. 2011, \aap, 530, A63

\bibitem[Peimbert(1967)]{peimbert67}
{Peimbert},~M. 1967, \apj, 150, 825

\bibitem[Rudy et al.(1985)]{rudy85}
{Rudy},~R.~J., {Stocke},~J.~T., \& {Foltz},~C.~B. 1985, \apj, 288, 531

\bibitem[Sadakane et al.(2010)]{sadakane10}
{Sadakane},~K., {Tajitsu},~A., {Mizoguchi},~S., {Arai},~A., \& {Naito},~H. 2010, \pasj, 62, L5 (Paper I)

\bibitem[Shore et al.(2011)]{shore11}
{Shore},~S.~N., {Augusteijn},~T., {Ederoclite},~A., \& {Uthas},~H. 2011, \aap, 533, L8

\bibitem[Shore et al.(2012)]{shore12}
{Shore},~S.~N., {Wahlgren},~G.~M., {Augusteijn},~T., {Liimets},~T., {Koubsky},~P., {{\v S}lechta},~M., \& {Votruba},~V. 2012, \aap, 540, A55

\bibitem[Simon et al.(2009)]{simon09}
{Simon},~J.~D, \etal\ 2009, \apj, 702, 1157

\bibitem[Williams et al.(2008)]{williams08}
{Williams},~R., {Mason},~E., {Della Valle},~M., \& {Ederoclite}, A. 2008, \apj, 685, 451

\bibitem[Williams \& Mason(2010)]{williams10}
{Williams},~R., \& {Mason},~E. 2010, \apss, 327, 207

\bibitem[Williams(2011)]{williams11}
{Williams},~R. 2011, astro-ph:arXiv1108.4917

\bibitem[Wilson et al.(1959)]{wilson59}
{Wilson},~O.~C., {Minich},~G., {Flather},~E., \& {Coffeen},~M.~F. 1959, \apjs, 4, 199

\bibitem[Yamaoka et al.(2007)]{yamaoka07}
{Yamaoka},~H., {Nakamura},~Y., {Nakano},~S., {Sakurai},~Y., \& {Kadota},~K. 2007, \iaucirc, 8803, 1

%\bibitem[Aauthor et al.(2001)]{key-1}
%  Aauthor, A., Bauthor, B., Cauthor, C.\ 2001, PASJ, vol, page
% Books
%\bibitem[Aauthor \& Author(2001a)]{key-2}
%  Aauthor, A., Author, B.\ 2001, Name of Book(Publisher, Tokyo) ch.0
% Books
%\bibitem[Aauthor \& Bauthor(2001b)]{key-3}
 % Aauthor, A., Bauthor, B.\ 2001, Name of Book(Publisher, Tokyo) page0
%......
% Editorial Books
%\bibitem[Dauthor(2001)]{key-n}
%  Dauthor A.~A.\ 2001, in Name of Book,
%   ed.\  D.~Editor (Publisher, Tokyo) page0
\end{thebibliography}
\end{document}